\RequirePackage{fix-cm}
\documentclass[twocolumn]{svjour3}          
\smartqed  
\usepackage{graphicx}
\journalname{Journal of Radioanalytical Nuclear Chemistry}
\begin{document}

\title{NEW DEVELOPMENTS IN THE EXPERIMENTAL DATA FOR CHARGED PARTICLE PRODUCTION OF MEDICAL RADIOISOTOPES
}


\author{F. Ditr\'oi \and F. T\'ark\'anyi \and S. Tak\'acs \and A. Hermanne
}


\institute{F. Ditr\'oi \and F. T\'ark\'anyi \and S. Tak\'acs \at
              Institute for Nuclear Research, Hungarian Academy of Sciences \\
              Tel.: +36-52-509251\\
              Fax: +36-52-416181\\
              \email{ditroi@atomki.hu}           
           \and
           A. Hermanne \at
              Cyclotron Laboratory, Vrije Universiteit Brussel (VUB), Brussels, Belgium     
}

\date{Received: 2014 / Accepted: 2014}

\maketitle

\begin{abstract}
The goal of the present work is to give a review of developments achieved experimentally in the field of nuclear data for medically important radioisotopes in the last three years. The availability and precision of production related nuclear data is continuously improved mainly experimentally. This review emphasizes a couple of larger fields: the Mo/Tc generator problem and the generator isotopes in general, heavy alpha-emitters and the rare-earth elements. Other results in the field of medical radioisotope production are also listed.

\end{abstract}

Keywords: charged particle induced nuclear reactions; medical radio-isotopes; review

\section{Introduction}
\label{intro}
Nuclear data play an important role in the production and the application of medical radioisotopes. Out of the basic nuclear data, the activation cross sections and the related production yields are also requested. These data can be obtained experimentally and/or by using nuclear reaction codes. The significance of the experimental data is twofold: they give direct information for the production and they also contribute to the development of nuclear reaction model codes. During the previous dedicated and satellite investigations, large amount of activation data were measured. In spite of the large experimental databases and the significant progress in the reaction theory, the importance of collection of reliable experimental data has not decreased. New candidate radioisotopes appear for diagnostics and therapy, the radioisotope impurities became more important, new experimental techniques appear to measure low radio-activities, weak performance of nuclear theory was recognized for some type of nuclear reactions and for the metastable states, preparation of recommended databases require re-measurement of mostly used reactions due to contradicting data. Application of high-energy accelerators, usefulness of deuteron induced reactions, broader application of numerous low energy cyclotrons all require further data measurements and improvements.
Nuclear data for medical applications are mainly used in the following fields: production of radioisotopes (medical), dosimetry calculations (staff, patients), radiotherapy planning, diagnostics and economy (planning of the production, distribution, treatment, etc.). 
The necessary nuclear data can be divided into two larger groups: i.e. structure and decay data, which are mainly interesting for fundamental research but also used for medical physics, and the excitation functions and production yields, which are rather application related data. The sources of stored nuclear data are manifold, if one is interested in the source publications, then turns to such publication databases as the WoS (Web of Science) \cite{1}. If more precise compiled data are necessary then the source is the EXFOR (Experimental Nuclear Reaction Data) \cite{2} database, and evaluated data can be found in the newest version of the ENDF (Evaluated Nuclear Data Library) \cite{3} libraries.

\section{Experimental data}
\label{sec:2}
The most important medically related radioisotopes and their basic nuclear data are listed in different literature sources \cite{4,5,6,7,8,9,10,11,12,13,14} and summarized in Table 1. These sources might not contain the new emerging radioisotopes under investigation, which are referred in the following sections. If really reliable production related data are needed, one can search in the experimental databases. The experimental groups perform their measurements by using: activation method, sometimes followed by chemical separation, on solid or gas targets. Single target, rotating wheel or stacked targets are used. The activity is measured by using $\gamma$-, X-ray and $\alpha$-spectrometry or mass spectrometry.
The most frequent goals of these studies are: new isotopes for medical, industrial and other purposes; new production routes of important isotopes; side reactions and impurities; clarification of disagreements between previous measurements.
According to our experiences, the published results are not always free of problems. These problems can be summarized as follows:
\begin{itemize}
\item	The available poor technology in some laboratories (beam intensity, beam energy measurement, target preparation, chemical separation)
\item	Missing basic knowledge on data measurement, data evaluation, definitions, etc.
\item	Too optimistic uncertainties 
\item	The reports are not detailed enough for data corrections and for production of covariance matrices
\end{itemize}
Some groups, especially those participating in coordinated research programs try to address these problems in the recent publications. In addition, people responsible for database inputs pay more attention to the above problems when evaluating a new work. Out of medical applications, the published results can also be used for basic nuclear physics research, development of theoretical models, improvement of computer codes based on these models, industrial applications, energy related and safety/security related applications. According to the last IAEA recommendations \cite{4} particular data improvements are needed at the following areas:
\begin{itemize}
\item	Monitor reactions (induced by light ions in the used energy ranges)
\item	Diagnostic and therapeutic $\gamma$-emitters ($^{123,121}$I, $^{123}$Cs, $^{123}$Xe, $^{51}$Cr, $^{186,188}$Re, $^{99m,g}$Tc, $^{90m,g}$Y, $^{99}$Mo)
\item	Positron emitters ($^{52}$Fe, $^{55}$Co, $^{61}$Cu, $^{66,68}$Ga, $^{90}$Nb, $^{89}$Zr, $^{73}$Se, $^{76}$Br, $^{86}$Y, $^{89}$Zr, $^{94m}$Tc, $^{110m}$In, $^{120}$I)
\item	Generators ($^{62}$Zn/$^{62}$Cu, $^{68}$Ge/$^{68}$Ga, $^{72}$Se/$^{72}$As,\\ $^{82}$Sr/$^{82}$Rb)
\item	Therapeutic $\alpha$-emitters ($^{225}$Ra, $^{225,227}$Ac, $^{230}$U, $^{227}$Th)
\item	Therapeutic electron and X-ray emitters ($^{131}$Cs, $^{178}$Ta)
\end{itemize}
Out of these recommendations also the $^{111}$In, $^{201}$Tl, $^{119}$Sb, $^{97}$Ru therapeutic electron and X-ray emitters require further data improvements. The recently emerged theranostic approach in nuclear medicine, which consists of performing imaging of the bio-distribution and therapy in the same time, by using the same isotope or different isotopes of the same element or isotopes of two elements with similar chemical behavior, opened a new field in nuclear data experiments and renewed the interest for some radionuclides. In the last 3-4 years, developments have been achieved in the field of experimental production data in the following four areas.

\begin{table*}[t]
\tiny
\caption{\textbf{List of important and emerging medically interesting radioisotopes, for which charged particle production route exist and/or measured recently \cite{1,2,3,4} (new measurements marked with *)}}
\label{tab:1}       
\centering
\begin{center}
\begin{tabular}{|p{0.8in}|p{0.9in}|p{1.5in}|p{1.9in}|p{1.5in}|} \hline 
\textbf{Radionuclide} & \textbf{Half-life} & \textbf{Decay mode} & \textbf{Reaction (main routes)} & \textbf{Application} \\ \hline 
${}^{11}$C* & 20.4 min & $\beta^{+}$ (100\%) & ${}^{14}$N(p,$\alpha$)${}^{11}$C & Positron emitter \\ \hline 
${}^{13}$N & 10 min & $\beta^{+}$ (100\%) & ${}^{16}$O(p,$\alpha$); ${}^{12}$C(d,n) & Radiotracer, PET, labelling \\ \hline 
${}^{15}$O* & 2.03 min & EC (0.1\%); $\beta^{+}$ (99.9\%) & ${}^{15}$N(p,n); ${}^{14}$N(d,n); ${}^{16}$O(p,pn); ${}^{12}$C($\alpha$,n) & Labelling, flow measurement \\ \hline 
${}^{18}$F* & 109.8 min & EC (3\%); $\beta^{+}$ (97\%) & ${}^{18}$O(p,n); \textbf{${}^{nat}$Ne(d,x)${}^{18}$F} & PET  \\ \hline 
${}^{22}$Na* & 2.6 a & EC (9.6\%); $\beta^{+}$ (90.6\%) & ${}^{22}$Ne(p,n); ${}^{22}$Ne(d,2n), ${}^{24}$Mg(d,$\alpha$) & PET calibration \\ \hline 
${}^{34m}$Cl & 32 min & IT (44.6\%); $\beta^{+}$ (54.3\%) & ${}^{nat}$Cl(p,pxn); ${}^{34}$S(p,n); ${}^{34}$S(d,2n) & PET \\ \hline 
${}^{38}$K* & ~7.636 min & EC (0.47\%); $\beta^{+}$ (99.53\%) & ${}^{35}$Cl($\alpha$,n), ${}^{38}$Ar(p,n) & PET \\ \hline 
${}^{43}$K* & 22.3 h & $\beta $${}^{-}$ 100~\% & ${}^{40}$Ar($\alpha$,p)${}^{43}$K & biology \\ \hline 
${}^{44}$Sc\newline \newline \textit{${}^{44}$Ti/${}^{44}$Sc*} & 3.97 h\newline \newline 59.1 a/ 3.97h & ~EC (5.73\%); $\beta^{+}$ (94.27\%)\newline \newline $\varepsilon $ (100\%)~ & ${}^{44}$Ca(p,n)${}^{44}$Sc; ${}^{44}$Ca(d,2n)${}^{44}$Sc; \newline ${}^{45}$Sc(p,2n)${}^{44}$Ti; ${}^{45}$Sc(d,3n)${}^{44}$Ti & PET \\ \hline 
${}^{51}$Cr* & 27.701 d & $\varepsilon $ (100\%) & ${}^{nat}$V(p,n);${}^{ nat}$V(d,2n);${}^{nat}$Ti($\alpha$,x)  & in vitro assay \newline $\gamma$-emitter \\ \hline 
${}^{52m}$Mn\newline \newline \textit{${}^{52}$Fe/${}^{52m}$Mn*} & 21.1 min\newline \newline 8.275~h & ~EC (3.25\%); $\beta^{+}$ (95.0\%); \newline IT (1.75\%)\newline \newline ~$\varepsilon $: (100\%) &  ${}^{55}$Mn(p,4n)${}^{52}$Fe; ${}^{nat}$Ni(p,x)${}^{52}$Fe; ${}^{52}$Cr(${}^{3}$He,3n)${}^{52}$Fe & PET \\ \hline 
${}^{55}$Co* & 17.6 h & EC (23\%); $\beta^{+}$ (77\%) & ${}^{56}$Fe(p,2n); ${}^{54}$Fe(d,n); ${}^{58}$Ni(p,$\alpha$), ${}^{nat}$Fe(p,x) & PET \\ \hline 
${}^{61}$Cu* & 3.333 h & EC (39\%); $\beta^{+}$ (61\%) & ${}^{61}$Ni(p,n); ${}^{nat}$Ni(d,x) & PET \\ \hline 
${}^{62}$Cu\newline \newline \textit{${}^{62}$Zn${}^{/62}$Cu}* & 9.67 min\newline \newline 9.193 h/~9.67 min\newline  & EC (2.17\%); $\beta^{+}$ (97.83 \%)\newline \newline $\varepsilon $ (100\%) & ${}^{63}$Cu(p,2n)${}^{62}$Zn; ${}^{63}$Cu(d,3n)${}^{62}$Zn; ${}^{60}$Ni($\alpha$,2n)  & PET \\ \hline 
${}^{64}$Cu* & 12.7 h & EC (44\%); $\beta^{+}$(17\%); $\beta^{-}$ (39\%) & ${}^{64}$Ni(p,n);${}^{ 64}$Ni(d,2n); ${}^{68}$Zn(p,$\alpha$n); ${}^{66}$Zn(d,$\alpha$) & PET, therapy, Cu metabolism \\ \hline 
${}^{67}$Cu* & 62 h & $\beta^{-}$ (100\%) & ${}^{64}$Ni($\alpha$,p); ${}^{68}$Zn(p,2p); ${}^{70}$Zn(p,$\alpha$) & Therapy \\ \hline 
${}^{63}$Zn & 38 min & EC (7\%); $\beta^{+}$ (93\%) & ${}^{63}$Cu(p,n) & PET biomarker for Zn \\ \hline 
${}^{66}$Ga* & 9.49 h & EC (43\%); $\beta^{+}$ (57 \%) & ${}^{66}$Zn(p,n); ${}^{63}$Cu($\alpha$,n) & PET \\ \hline 
${}^{67}$Ga* & 3.26 d & EC (100\%) & ${}^{67}$Zn(p,n); ${}^{68}$Zn(p,2n) & SPECT \\ \hline 
${}^{68}$Ga*\newline \textit{${}^{68}$Ge/${}^{68}$Ga*} & 68 min\newline 270.8 d/68 min & EC (11\%); $\beta^{+}$ (89\%)\newline $\varepsilon $ (100\%) & ${}^{68}$Zn(p,n)\newline ${}^{nat}$Zn($\alpha$,x); ${}^{nat}$Ga(p,x); ${}^{69}$Ga(p,2n) & PET imaging, PET calibration \\ \hline 
${}^{72}$As*\newline \textit{${}^{72}$Se/${}^{72}$As} & 26.0 h\newline 8.40 d/26.0 h & EC (12.2\%); $\beta^{+}$ (87.8\%)\newline ${\rm E}$ (100\%) & ${}^{nat}$Ge(p,xn)${}^{72}$As\newline ${}^{75}$As(p,4n)${}^{72}$Se; ${}^{ nat}$Br(p,x)${}^{72}$Se & PET \\ \hline 
${}^{73}$As* & 8.3 d & $\varepsilon $ (100\%) & ${}^{nat}$Ge(p;$\alpha$,x) & PET and labelling \\ \hline 
${}^{74}$As* & 17.8 d & $\varepsilon $ (100\%) & ${}^{74}$Ge(d;p,xn); ${}^{nat}$Ga($\alpha$,x) & PET (cancer diagnostics) \\ \hline 
${}^{73}$Se* & 39.8 min & ~EC (79.6\%); $\beta^{+}$ (20.4\%) & ${}^{75}$As(p,3n), ${}^{72}$Ge($\alpha$,3n) & PET \\ \hline 
${}^{75}$Br* & 97 min & $\beta^{+}$ (75\%) EC (25\%) & ${}^{7x}$Se(p;d;${}^{3}$He;$\alpha$,X); ${}^{78}$Kr(p,$\alpha$) & PET \\ \hline 
${}^{76}$Br* & 16.2 h & $\beta^{+}$ (54\%) EC (46\%) & ${}^{76}$Se(p,n); ${}^{77}$Se(p,2n); ${}^{75}$As($\alpha$,3n), \newline ${}^{nat}$Se(p,xn); ${}^{nat}$Br(p,xn); ${}^{nat}$Br(d,xn), ${}^{78}$Kr(d,$\alpha$) & PET \\ \hline 
${}^{77}$Br & 57 h & EC (99.3\%)~$\beta^{+}$(0.73 \%) & ${}^{nat}$Se(p,xn); ${}^{75}$As($\alpha$,2n) & SPECT, therapy \\ \hline 
${}^{81m}$Kr\newline \textit{${}^{81}$Rb/${}^{81m}$Kr*} & 13 s\newline 4.6 h/13 s & IT (99.9975\%) & ${}^{81}$Rb daughter\newline ${}^{82}$Kr(p,2n); ${}^{80}$Kr(d,n) & Pulmonology, SPECT \\ \hline 
${}^{82}$Rb\newline \textit{${}^{82}$Sr/${}^{82}$Rb*} & 1.3 min\newline 25 d/1.3 min & EC (4.57\%); $\beta^{+}$ (95.43\%)\newline $\varepsilon $ (100\%) & ${}^{82}$Sr daughter\newline ${}^{85}$Rb(p,4n); ${}^{nat}$Rb(p,xn); ${}^{82}$Kr($\alpha$,4n), ${}^{82}$Kr(${}^{3}$He,3n) & PET \\ \hline 
${}^{86}$Y* & 14.7 h & EC (66\%); $\beta^{+}$ (34\%) & ${}^{86}$Sr(p,n); ${}^{88}$Sr(p,3n); ${}^{nat}$Zr(d,x) & Tracing of ${}^{90}$Y bone pain palliation agent \\ \hline 
${}^{90}$Nb* & 14.60 h & EC (48.8\%); $\beta^{+}$ (51.2\%) & ${}^{90}$Zr(p,n); ${}^{90}$Zr(d,2n) & PET \\ \hline 
${}^{88}$Y\newline \textit{${}^{88}$Zr/${}^{88}$Y*} & 106 d\newline 83.4 d/106 d & $\varepsilon $ (100\%)\newline $\varepsilon $ (100\%) & ${}^{88}$Sr(p,n)${}^{ 88}$Y; ~${}^{nat}$Rb($\alpha $,xn)${}^{88}$Y\newline  ${}^{89}$Y(p,2n); ${}^{89}$Y(d,3n) & Tracing of ${}^{90}$Y \\ \hline 
${}^{89}$Zr* & 3.3 d & EC (77\%); $\beta^{+}$ (23\%) & ${}^{89}$Y(p,n); ${}^{89}$Y(d,2n) & PET, labelling \\ \hline 
${}^{94m}$Tc\newline  & 52 min & EC (28\%); $\beta^{+}$ (72\%) & ${}^{nat}$Mo(p,n); ${}^{nat}$Mo($\alpha$,x) also on enriched targets & PET imaging, ${}^{99m}$Tc replacement \\ \hline 
${}^{99m}$Tc*\newline \textit{${}^{99}$Mo/${}^{99m}$Tc*} & 6.0067 h \newline 65.976 /6.0067 h  & IT (99.9963\%); $\beta^{-}$ (0.0037\%)\newline $\beta $${}^{-}$ (100\%)~ & ${}^{100}$Mo(p,2n)${}^{99m}$Tc;${}^{ 100}$Mo(d,3n)${}^{99m}$Tc\newline ${}^{100}$Mo(p,pn)${}^{99}$Mo; ${}^{100}$Mo(d,p2n)${}^{99}$Mo & SPECT \\ \hline 
${}^{103}$Pd* & 17 d & $\varepsilon $ (100\%) & ${}^{103}$Rh(p,n); ${}^{103}$Rh (d,2n) & Brachytherapy \\ \hline 
${}^{111}$Ag* & 7.45 d & $\beta $${}^{-}$ (100\%)~ & ${}^{110}$Pd(d,n) & Therapy \\ \hline 
${}^{109}$Cd* & 461.4 d & EC (100\%) & ${}^{109}$Ag(p,n); ${}^{107}$Ag($\alpha$,x) & ${}^{109}$Cd/${}^{109m}$Ag biomedical generator \\ \hline 
${}^{110m}$In*\newline \textit{${}^{110}$Sn/${}^{110m}$In*} & 69 min\newline 4.9 h/69 min & EC (99\%); $\beta^{+}$ (0.008\%)\newline $\varepsilon $ (100\%) & ${}^{110}$Cd(p,n); ${}^{110}$Cd(d,2n)\newline ${}^{nat}$In(p,x); ${}^{nat}$Cd(${}^{3}$He,x); ${}^{nat}$Cd($\alpha$,x) & PET analogue of 111In; labelling \\ \hline 
${}^{111}$In* & 2.83 d & EC (100\%) & ${}^{111}$Cd(p,n); ${}^{112}$Cd(p,2n); ${}^{109}$Ag($\alpha$,2n) & SPECT, diagnostics \\ \hline 
${}^{113m}$In*\newline \newline \textit{${}^{113}$Sn/${}^{113m}$In*} & 99.476 min\newline \newline 115.09 d~ & IT (100\%)\newline \newline EC (100\%) & ${}^{nat}$Cd(p,x);${}^{ 114}$Cd(p,2n);${}^{ nat}$Cd(d,x)${}^{113m}$In; ${}^{114}$Cd(d,3n)${}^{113m}$In\newline ${}^{nat}$Cd(${}^{3}$He,xn)${}^{113}$Sn; ${}^{nat}$Sn(p,x)${}^{113}$Sn; ${}^{nat}$Sn(d,x)${}^{113}$Sn; ${}^{nat}$Cd($\alpha $,xn)${}^{113}$Sn; ${}^{113}$In(p,x)${}^{113}$Sn; ${}^{nat}$In(p,x)${}^{113}$Sn; ${}^{113}$In(d,x)${}^{113}$Sn;${}^{ nat}$In(d,x)${}^{113}$Sn; ${}^{1}$${}^{11}$Cd($\alpha $,x)${}^{113}$Sn & SPECT, radio-tracer, therapy \\ \hline 
${}^{114m}$In* & 49.5 d & IT (96.75\%); e (3.25\%) & ${}^{114}$Cd(p,n); ${}^{116}$Cd(p,3n); ${}^{114}$Cd(d,2n) & therapy, radio-tracer \\ \hline 
${}^{120g}$I & 1.35 h & EC (31.8\%); $\beta^{+}$ (68.2\%) & ${}^{nat}$Te(p,xn) & PET \\ \hline 
${}^{117m}$Sn* & ~14.00 d & IT (100\%) & ${}^{116}$Cd($\alpha$,3n); ${}^{116}$Cd(${}^{3}$He,2n) & therapy \\ \hline 
${}^{121}$I & 2.12 h & EC (91.4\%); $\beta^{+}$ (10.6\%) & ${}^{122}$Te(p,2n) & SPECT \\ \hline 
${}^{122}$I*\newline \textit{${}^{122}$Xe/${}^{122}$I*} & 3.63 min\newline 20.1 h/3.63 min & EC (22\%); $\beta^{+}$ (78\%)\newline $\varepsilon $ (100\%) & ${}^{122}$Te(d,2n)\newline ${}^{127}$I(p,6n); ${}^{127}$I(d,7n); ${}^{124}$Xe(p,x) & PET analogue of ${}^{123,125,131}$I \\ \hline 
${}^{123}$I* & 13.2 h & EC (100\%) & ${}^{123}$Te(p,n); ${}^{124}$Te(p,2n); ${}^{122}$Te(d,n); ${}^{124}$Xe(p,x) & Thyroid SPECT diagnostics \\ \hline 
${}^{124}$I* & 4.2 d & EC (78.3\%); $\beta^{+}$ (22.7\%) & ${}^{123}$Te(d,n); ${}^{124}$Te(p,n); ${}^{124}$Te(d,2n), ${}^{121}$Sb($\alpha$,n) & PET, diagnostic and therapy \\ \hline 
${}^{127}$Xe & 34.4 d & EC (100\%) & ${}^{127}$I(p,n) & SPECT \\ \hline 
${}^{128}$Cs\newline \textit{${}^{128}$Ba${}^{/128}$Cs*} & 3.62 min\newline 2.43 d/3.62 min & EC (31.1\%); $\beta^{+}$ (68.9\%) ~\newline $\varepsilon $ (100\%) &  ${}^{133}$Cs(p,5n)${}^{128}$Ba & PET analogue of ${}^{131}$Cs\newline Potassium analogue \\ \hline 
${}^{131}$Cs*\newline \textit{${}^{131}$Ba/${}^{131}$Cs*} & 9.689 d\newline 11.50 d/9.689 d & EC (100\%)\newline $\varepsilon $ (100\%) & ${}^{131}$Xe(p,n)\newline ${}^{133}$Cs(p,3n)${}^{131}$Ba $\rightarrow$ ${}^{131}$Cs & Brachytherapy \\ \hline 
${}^{145}$Sm*\newline \textit{${}^{145}$Eu/${}^{145}$Sm*} & 340 d\newline 5.93 d & $\varepsilon $ (100\%)\newline EC (98.099\%); $\beta^{+}$ (1.91\%)  & ${}^{144}$Sm(d,x)${}^{145}$Sm\newline ${}^{nat}$Sm(p,x)${}^{145}$Eu $\rightarrow$${}^{145}$Sm & Brachytherapy~\newline Therapy \\ \hline 

\end{tabular}
\end{center}

\end{table*}

\setcounter{table}{0}
\begin{table*}[t]
\tiny
\caption{ cont.}
\label{ cont.}
\centering
\begin{center}
\begin{tabular}{|p{0.8in}|p{0.9in}|p{1.5in}|p{1.9in}|p{1.5in}|} \hline 
\textbf{Radionuclide} & \textbf{Half-life} & \textbf{Decay mode} & \textbf{Reaction (main routes)} & \textbf{Application} \\ \hline 
${}^{153}$Sm* & 46.50 h & $\beta $${}^{-}$ (100\%) & ${}^{152}$Sm(d,p) & Therapy \\ \hline 
${}^{149}$Pm & 53.08 & $\beta $${}^{-}$: (100\%) & ${}^{148}$Nd(d,x); ${}^{150}$Nd(d,x);${}^{ 150}$Nd${}^{ }$(p,x) & Therapy \\ \hline 
${}^{140}$Nd/${}^{140}$Pr* & ~~3.37 d\newline ~3.39 min & $\varepsilon $ (100\%)~\newline EC (49\%); $\beta^{+}$ (51.0\%)  \newline  & ${}^{141}$Pr(p,2n); ${}^{141}$Pr(d,2n); ${}^{nat}$Nd(d,x), ${}^{nat}$Nd(p,x) & ${}^{140}$Pr therapy, ${}^{140}$Nd PET \\ \hline 
${}^{161}$Tb* & 6.89 d & $\beta $${}^{-}$ (100\%) & ${}^{160}$Gd(d,n) & Therapy \\ \hline 
${}^{161}$Ho* & ~2.48 h & $\varepsilon $ (100\%) & ${}^{161}$Dy(p,n)${}^{161}$Ho${}^{, 162}$Dy (p,2n); ${}^{161}$Dy(d,2n) & Therapy \\ \hline 
${}^{165}$Er* & 10.36 h & $\varepsilon $ (100\%)~ & ${}^{165}$Ho(p,n);\textit{~}${}^{165}$Ho(d,2n); ${}^{nat}$Er(p,x)${}^{165}$Tm $\rightarrow$${}^{165}$Er, ${}^{nat}$Er(d,x)${}^{165}$Tm$\rightarrow$${}^{165}$Er\textit{} & Therapy \\ \hline 
${}^{167}$Tm* & 9.25 d & $\varepsilon $ (100~\%)~ & ${}^{165}$Ho($\alpha$,2n); ${}^{167}$Er(p,n); ${}^{167}$Er(d,2n); ${}^{nat}$Yb(p,xn)${}^{167}$Lu${}^{167}$Yb${}^{167}$Tm\newline  & Therapy, SPECT \\ \hline 
${}^{169}$Yb* & 32.018 d & $\varepsilon $ (100~\%)~ & ${}^{169}$Tm(p,n); ${}^{169Tm}$(d,2n);${}^{ 168}$Er($\alpha$,2n) & Therapy \\ \hline 
${}^{177}$Lu* & 6.71 d & $\beta^{-}$ (100\%) & ${}^{176}$Yb(d,x) & Therapy  \\ \hline 
${}^{178}$Ta*\newline \textit{${}^{178}$W/${}^{178}$Ta*\newline } & 2.36 h\newline 21.6 d/2.36 h\newline  & EC (100\%)\newline $\varepsilon $ (100\%) & ${}^{nat}$Hf(p,x)${}^{178}$Ta\newline ${}^{nat}$Ta(p,x)${}^{178}$W; ${}^{nat}$Ta(d,x)${}^{178}$W; ${}^{178}$Hf($\alpha$,2n)${}^{ 178}$W\newline \newline  & Therapy (e${}^{-}$, X-ray) \\ \hline 
${}^{186}$Re* & 3.72 d & $\beta $${}^{-}$ (92.53\%), $\varepsilon $  (7.47\%) & ${}^{186}$W(d,2n); ${}^{186}$W(p,n) & Therapy \\ \hline 
${}^{192}$Ir * & 73.829 d & $\beta $${}^{-}$ (95.24\%), $\varepsilon $ (4.76~\%) & ${}^{192}$Os(p,n); ${}^{192}$Os(d,2n) & Brachytherapy \\ \hline 
${}^{191}$Pt* & 2.802~d~ & EC (100\%) & ${}^{193}$Pt(p,3n) & Diagnostic for therapy \\ \hline 
${}^{198}$Au* & ~2.6947 d & $\beta $${}^{-}$ (100\%) & ${}^{198}$Pt(p,n) & Brachytherapy \\ \hline 
${}^{199}$Au* & ~3.139 d & $\beta $${}^{-}$ (100\%) & ${}^{198}$Pt(d,n) & Therapy \\ \hline 
${}^{195}$Au\newline \textit{${}^{195m}$Hg/${}^{195m}$Au*} & 30.5 s\newline 41 h/30.5 s & IT (100\%)\newline EC (45.8\%); IT (54.2\%) & ${}^{197}$Au(p,3n); ${}^{nat}$Pt($\alpha$,x) & Angiocardiography \\ \hline 
${}^{201}$Tl\newline \textit{${}^{201}$Pb/${}^{201}$Tl*} & 73.1 h\newline \newline 9.33 h & EC (100\%)\newline \newline EC (99.936\%); $\beta^{+}$ (0.064\%) & ${}^{203}$Tl(p,3n)${}^{201}$Pb  ${}^{201}$Tl & SPECT, myocardia \\ \hline 
${}^{203}$Pb & 51.9 h & $\varepsilon $ (100\%) & ${}^{nat}$Tl(p,x); ${}^{205}$Tl(p,3n) & In vivo, in vitro studies \\ \hline 
${}^{225}$Ac ${}^{ }$and ${}^{213}$Bi \newline ${}^{229}$Th${}^{-223}$Ra decay chain\newline ${}^{233}$U($\alpha$) ${}^{229}$Th($\alpha$)${}^{225}$Ra($\beta^{-}$) ${}^{225}$Ac($\alpha$) ${}^{221}$Fr($\alpha$)\newline ${}^{217}$At($\alpha$)${}^{213}$Bi($\beta^{-}$)\newline ${}^{213}$Po($\alpha$)${}^{209}$Pb($\beta^{-}$) ${}^{209}$Bi(stable)\newline  & 10.0 d/60.55 min & $\beta^{- }$(97.8\%), a (2.2\%) & ${}^{226}$Ra(p,2n${}^{)225}$Ac; ${}^{232}$Th(p,x)${}^{225}$Ac & $\alpha$-immunotherapy\newline $\alpha$ targeted therapy \\ \hline 
${}^{211}$At & 7.2 h & EC (59\%), a (41\%) & ${}^{209}$Bi($\alpha$,2n) & $\alpha$ targeted therapy \\ \hline 
${}^{227}$Th-${}^{223}$Ra decay chain\newline \textbf{${}^{227}$Th($\alpha$)${}^{223}$Ra($\alpha$)}\newline ${}^{219}$Rn($\alpha$) ${}^{215}$Po($\alpha$)\newline ${}^{211}$Pb($\beta^{-}$) ${}^{211}$Bi${}^{207}$Tl($\beta^{-}$)\newline ${}^{207}$Pb(stable) & 18.68 d/11.43 d & $\alpha$ (100\%)\newline $\alpha $ (100\%) & ${}^{nat}$Th(p,x) & $\alpha$ targeted therapy \\ \hline 
${}^{230}$U-${}^{226}$Th decay chain\newline \textbf{${}^{230}$U(${\mathbf \alpha }$)${}^{226}$Th(${\mathbf \alpha }$)}\newline ${}^{222}$Ra($\alpha $)${}^{218}$Rn($\alpha $)\newline ${}^{214}$Po($\alpha $)${}^{210}$Pb($\beta $${}^{-}$)\newline ${}^{210}$Bi($\beta $${}^{-}$)${}^{210}$Po($\alpha $) ${}^{206}$Pb(stable) & ~20.8 d/~30.57 min & $\alpha $ (100\%)\newline $\alpha $ (100\%) & ${}^{231}$Pa(p,2n)${}^{230}$U & $\alpha$ targeted therapy \\ \hline 
\end{tabular}

\end{center}
\end{table*}

\subsection{Production of $^{99}$Mo/$^{99m}$Tc generator}
\label{2.1}

Shutdown and decommissioning of several nuclear reactors responsible for the large-scale production of $^{99}$Mo caused (or will cause in the near future) a shortage on the market of the $^{99}$Mo/$^{99m}$Tc generator system. Several attempts have been made to elaborate methods to substitute the reactor HEU (highly enriched uranium) production route by LEU (low enriched uranium target), fast neutron \cite{15,16}, high energy photon \cite{17} and charged particle irradiation, both for $^{99}$Mo production \cite{18,19} and direct $^{99m}$Tc production \cite{20}.  Fig. 1 shows results on charged particle production routes for proton, deuteron and $\alpha$-particle induced reactions on natural and enriched molybdenum targets.
From Fig. 1, one can see that if there are more than one published experimental datasets for a reaction type (e.g. (d,x) reaction by Chodash \cite{18} and T\'ark\'anyi \cite{6}) ), the agreement between the independent experiments is very good. If the results on enriched targets are normalized to be compared with natural targets, the agreement is also good except for the lower energy part of the measured excitation functions. Fig. 1 also shows that the alternative production route of $^{99}$Mo can be the deuteron activation, but the availability of higher energy deuterons ($E_d \>$ 15 MeV), by the commercial medical accelerators, is limited. Moreover, deuterons have a shorter range in matter than protons, which gives a lower thick target yield for equivalent cross section values, that's why the proton production route is also competitive \cite{21}.

\begin{figure}
  \includegraphics[width=0.5\textwidth]{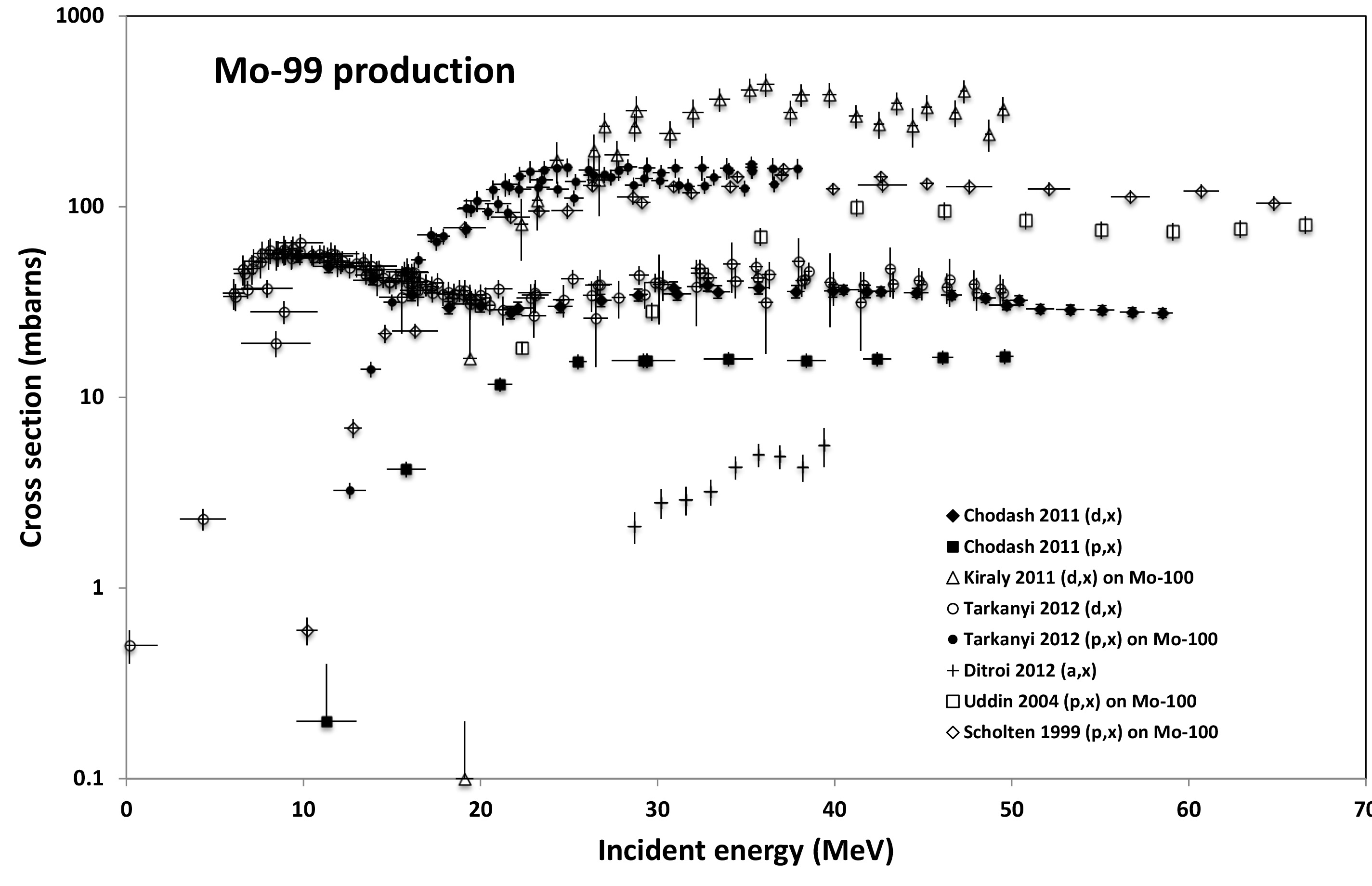}
\caption{Recent results on $^{99}$Mo production by charged particle activation on natural Mo \cite{6,7,18,19} and $^{100}$Mo enriched targets \cite{20} }
\label{fig:1}       
\end{figure}

\subsection{Heavy $\alpha$-emitters}
\label{2.2}

$\alpha$-emitter isotopes have a promising future in the field of targeted radio-therapy. The emitted $\alpha$-particles are relatively heavy projectiles, having different energies with short range in human tissues. That's why they can easily induce the medically required double-strand DNA break in the tumor cells without damaging the surrounding healthy cells. New results have been published for the $\alpha$-particle and proton induced production of $^{211}$At \cite{22} and $^{225}$Ac \cite{23,24} respectively.  The actinide radioisotope $^{225}$Ac has a half-life of 10 days, emits four alpha particles in its decay chain, and has recently gained importance for application in future treatment of metastatic cancer via targeted $\alpha$-immunotherapy \cite{25}. $^{225}$Ac can also be used as a generator for $^{213}$Bi, another shorter-lived $\alpha$-emitter ($T_{1/2}$ = 45.6 min) considered for targeted alpha therapy. Up till now, the widespread use of $^{225}$Ac and $^{213}$Bi in radiotherapy has been restricted by the limited availability of $^{225}$Ac. Presently, $^{225}$Ac is almost exclusively supplied by separating the isotope radio-chemically from only two $^{229}$Th sources, one located at Oak Ridge National Laboratory (ORNL), USA \cite{26} and the other at the Institute for Transuranium Elements in Karlsruhe (ITU), Germany \cite{27}. The $^{229}$Th, available at both sites was recovered from $^{233}$U, which was in long-term storage at ORNL. This $^{233}$U was produced in kilogram quantities in the 1960's by neutron irradiation of $^{232}$Th in molten salt breeder reactors \cite{28}.
Another radionuclide of interest, $^{223}$Ra, has a half-life of 11.4 days and also emits four alpha particles in its decay chain. It is a promising candidate for the treatment of bone cancer \cite{29}. It can also be used as a generator for the production of $^{211}$Pb \cite{30}. Proton induced production of more than 80 isotopes on thorium was measured at high and low proton energies \cite{31,32}. New results on production of $^{223}$Ra, $^{225}$Ac \cite{24} and $^{227}$Th \cite{23,24} were published recently. As an example, in Fig. 2, recent results (2012) for proton-induced reaction cross section of $^{225}$Ac on a thorium target are shown. Two groups measured this excitation function in overlapping energy ranges. The agreement in the medium energy region is not so good. 

\begin{figure}
  \includegraphics[width=0.5\textwidth]{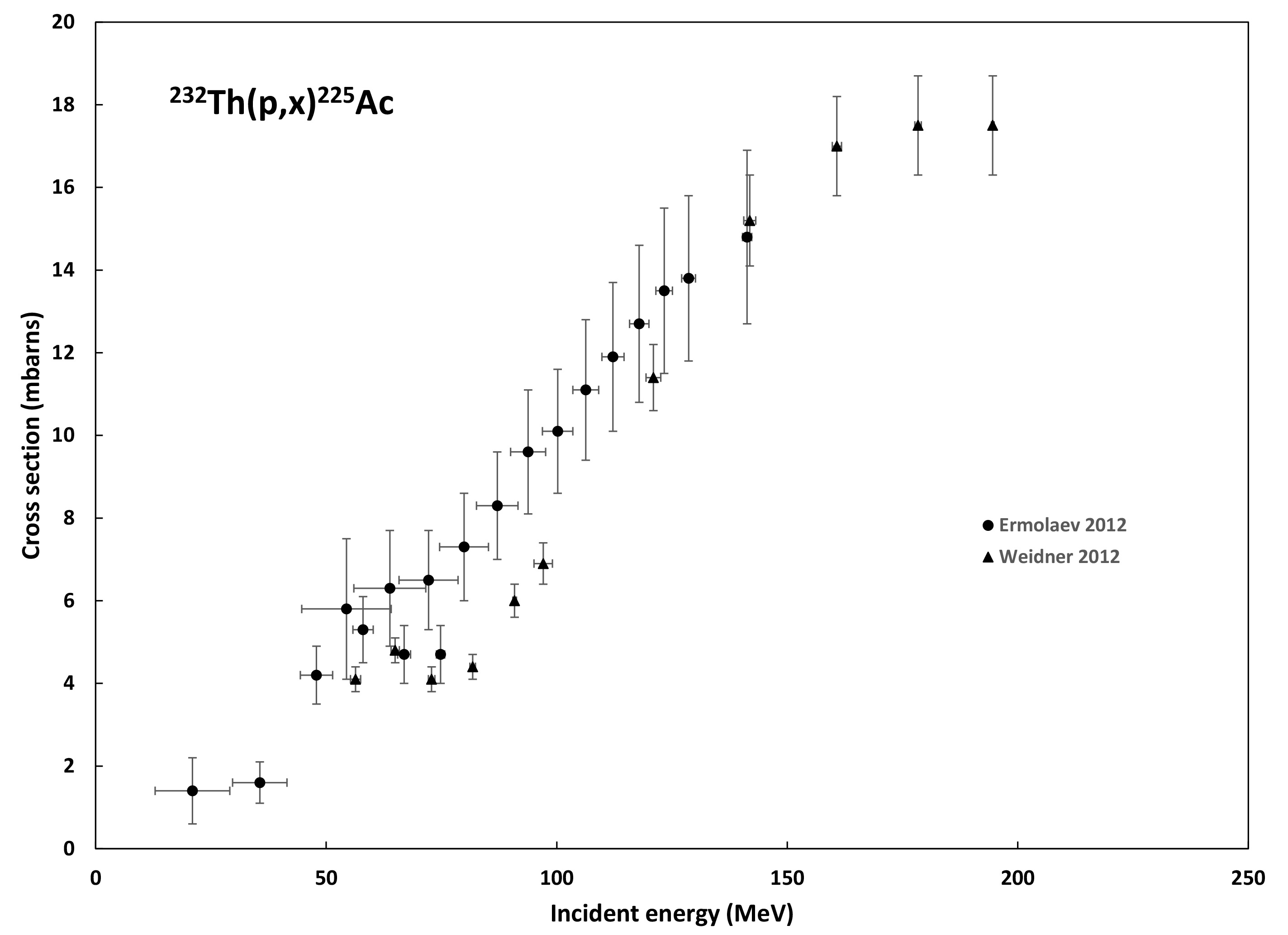}
\caption{New results on the production of $^{225}$Ac by high energy proton irradiation of $^{232}$Th}
\label{fig:2}       
\end{figure}

\subsection{Rare earth elements}
\label{2.3}

As there are a lot of emerging medical radioisotopes between the rare earth elements, most of the new results have been published in this field. Out of the rare earth radioisotopes, also new data for other important radio-products (side products) were determined. New experimental data were published for proton induced production of Hf, Lu and Ta radioisotopes on natural Hf target \cite{33}, proton induced production of Ho radioisotopes on Dy target \cite{34}, Zr, Y, Sr, Rb radioisotope production on natural yttrium by proton irradiation \cite{35}, measurement of the excitation functions of Pm, Nd, Pr and Ce radioisotopes on natural Nd target with deuteron activation \cite{36}, production of Yb and Tm radioisotopes on thulium target by proton activation \cite{37}. As an example, the new production cross section of the medically interesting therapeutic $^{161}$Ho radio-lanthanide, which is an Auger-electron emitter, is presented in Fig. 3. This cross section has not been measured before, that's why the results were compared with the output of the different theoretical nuclear reaction model codes (such as TALYS, ALICE and EMPIRE \cite{38,39,40}). These codes give good trend but different and incorrect quantitative estimation for the cross section. 

\begin{figure}
  \includegraphics[width=0.5\textwidth]{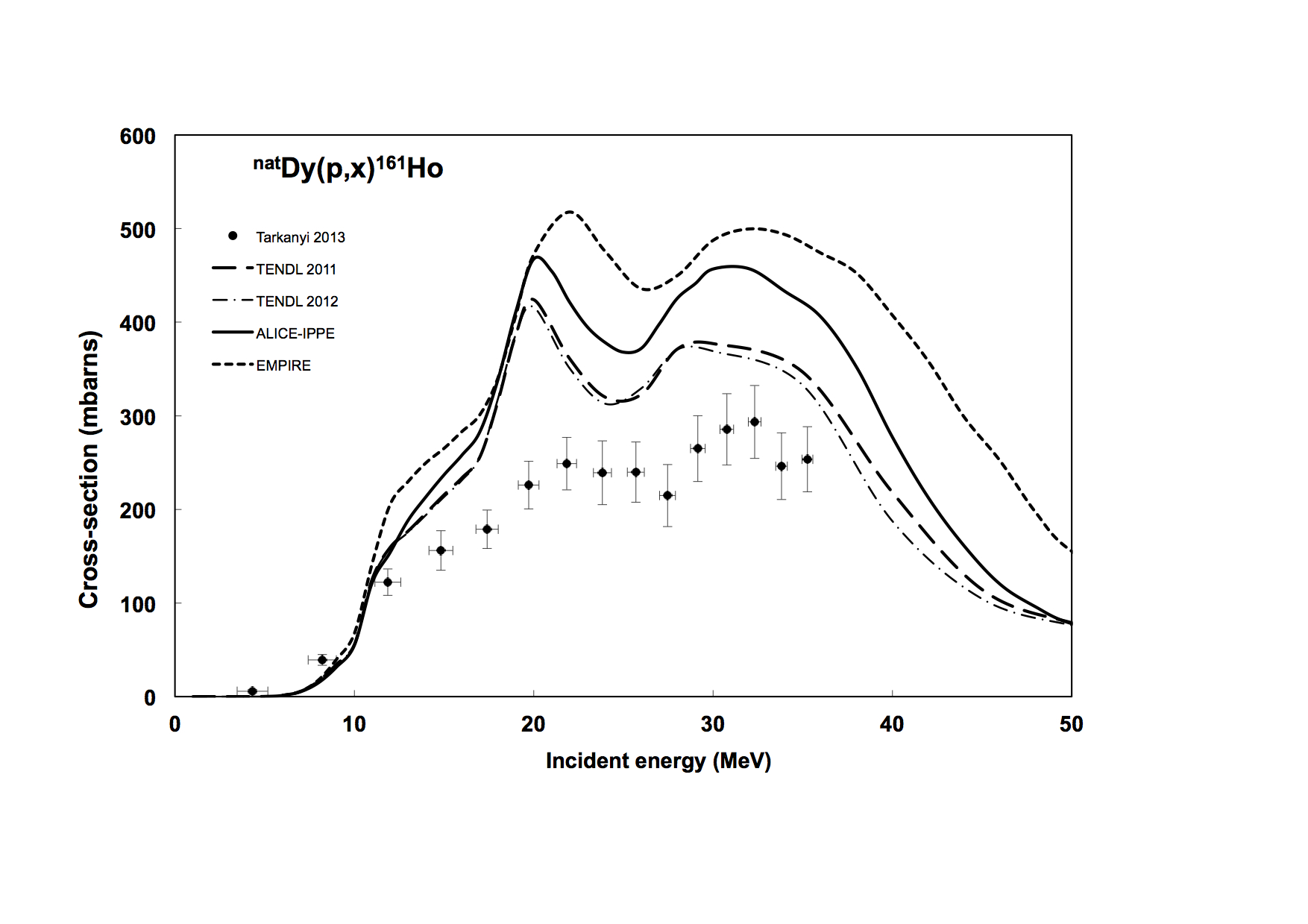}
\caption{Excitation function of the medically important $^{161}$Ho radioisotope by proton bombardment of natural dysprosium and comparison with the results of the different nuclear reaction model codes}
\label{fig:3}       
\end{figure}

\subsection{Generators}
\label{2.4}

Radioisotope generators are very important in the everyday medical practice, because they make possible the access to short-lived radioisotopes without the presence of a medical accelerator in the vicinity of the hospital. The most important generator/daughter pair is the $^{99}$Mo/$^{99m}$Tc, which was discussed in detail in a previous section.  As an example for this part, the new results on $^{62}$Zn for $^{62}$Cu generator is shown in Fig. 4. It contains both proton \cite{32,41,42,43} and deuteron \cite{44} induced routes. From the results presented in Fig. 4, only the data of Jost \cite{32} for proton irradiation and the data of Simeckova \cite{44} for deuteron irradiation are new (from the last three years), the others are presented for comparison. The figure shows that the newest $^{62}$Zn data for proton irradiation support the previous results and also confirm that the proton induced route is preferable than the deuteron one for its production. New results have also been published for $^{178}$W production (generator for $^{178}$Ta) \cite{45,46}, for $^{44}$Ti production (generator for $^{44}$Sc) \cite{47}, for $^{140}$Nd (generator for $^{140}$Pr) \cite{48} and for $^{149}$Pm (generator of $^{149}$Nd) \cite{36}. 

\begin{figure}
  \includegraphics[width=0.5\textwidth]{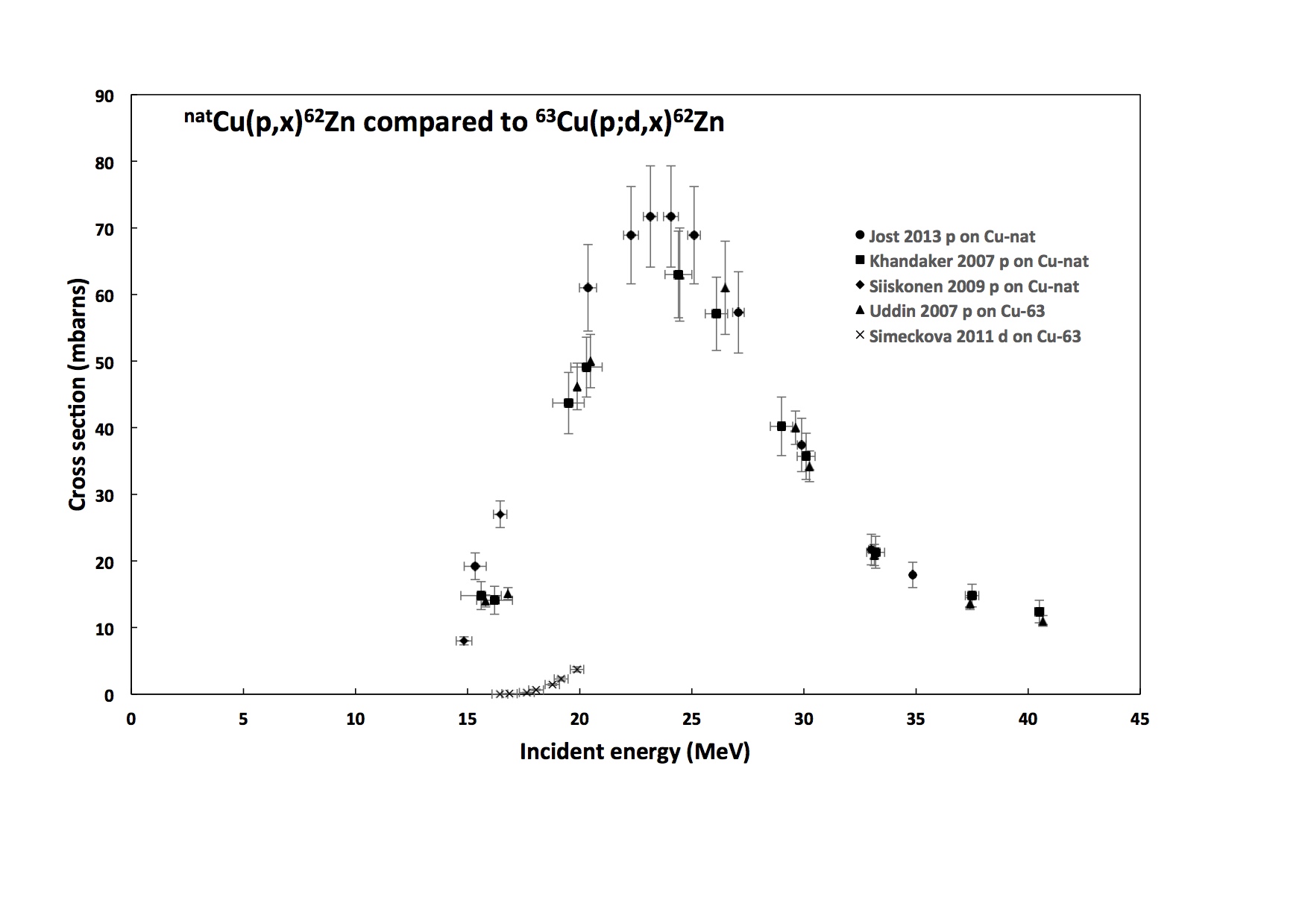}
\caption{Latest results on production cross sections for $^{62}$Zn with proton (Jost, Khandaker, Siiskonen, Uddin \cite{32,41,42,43}) and deuteron (Simeckova \cite{44}) irradiation of copper targets}
\label{fig:4}       
\end{figure}

\subsection{Other new results}
\label{2.5}

New results on other medically interesting nuclear data have also been published in the last three years: 
\begin{itemize}
\item	Production cross section of $^{110m,110g,111g,114m}$In on natural cadmium with $\alpha$-particle irradiation and many side reactions producing In, Cd and Sn radioisotopes \cite{49} (PET, electron emitter respectively).
\item	Production cross section of $^{186g}$Re with deuteron \cite{50} and proton \cite{45} irradiation on natural tungsten target and side products (therapy).
\item	Production cross section of $^{203,201}$Pb and $^{201}$Tl on natural thallium with proton irradiation and many side products \cite{51} (SPECT).
\item	Production parameters of $^{105}$Rh by proton irradiation of natural palladium \cite{52} (therapy).
\item	Production cross section of $^{167}$Tm and many side thulium and ytterbium products by proton irradiation of natural thulium \cite{37} (therapy).
\item	Production cross section of $^{66,67,68}$Ga by proton irradiation of $^{68}$Zn and $\alpha$-particle irradiation of natural copper \cite{53} (SPECT, PET).
\item	Production cross section of $^{90,95g}$Nb and $^{88}$Y with side products by proton irradiation of natural zirconium \cite{54} (PET).
\item	Production cross section of $^{11}$C, $^{15}$O and $^{13}$N isotopes by proton irradiation of the corresponding target elements \cite{55} (PET).
\item	Production of $^{67}$Cu by deuteron irradiation of natural zinc \cite{56} (therapy).
\item	Production cross section $^{123}$Cs by proton irradiation of $^{124}$Xe \cite{57} (SPECT).
\item	Production cross section of $^{123}$Xe by $\alpha$-particle \cite{58} and proton irradiation of $^{120}$Te and $^{124}$Xe targets \cite{57} respectively (SPECT).
\item	Production cross section of $^{51}$Cr by proton irradiation of natural chromium, $^{56}$Fe \cite{59}, natural nickel, $^{93}$Nb \cite{60}, $^{55}$Mn \cite{61} and $^{59}$Co \cite{62} as well as by deuteron irradiation of natural iron \cite{63,64}, natural vanadium \cite{65}, $^{55}$Mn \cite{66}, natural chromium \cite{67} and natural nickel \cite{68,69}, production of $^{52}$Mn by proton irradiation of natural chromium \cite{70} (SPECT, PET). 
\item	Production cross section of $^{90}$Nb by proton induced reactions of natural zirconium \cite{54}, $^{93}$Nb \cite{60,5} and natural molybdenum \cite{19}, deuteron \cite{6} and $\alpha$-particle \cite{7} induced reaction on natural molybdenum (PET).
\item	Production cross sections of $^{147,149}$Gd by proton and deuteron induced reactions on natural europium \cite{71} (SPECT).
 \end{itemize}

\section{Conclusions}
\label{sec:3}
The status of the nuclear reaction databases for medical applications has been significantly improved. Further developments are important in both the reaction and decay fields. It requires new, dedicated experiments, reliable (proved) experimental technique, new evaluation technique and proper reporting of the experimental data. 
Some isotopes are emerging, especially in the field of theranostic applications. It requires further measurements for these new isotopes and also re-measurements of isotopes having renewed interest. The results provided by the nuclear reaction model codes are not always satisfactory, that's why further and closer co-operation is necessary between the code developers and experimental groups. 
The huge amount of new data on medical radioisotopes confirms the large potential of experimental groups behind the work. The IAEA NDS is only one organization trying to promote and improve the co-operation in the nuclear data, especially the medical nuclear data field. Independent groups also provided important contribution to the experimental nuclear data for the medical radioisotopes (see in the section "Other new results").

\clearpage
\bibliographystyle{spphys}       
\bibliography{ici8}   

%
%

\end{document}